\documentclass[11pt]{article}
\usepackage[T2A]{fontenc}
\usepackage[utf8]{inputenc}
\usepackage[english]{babel}
\usepackage{graphicx}
\usepackage{wrapfig}
\usepackage{bmpsize}
\graphicspath{{Images/}}
\usepackage{amsmath,amsfonts,amssymb,amsthm,mathtools}
\usepackage{xcolor}
\usepackage{subcaption}
\usepackage{color}
\usepackage{indentfirst}

\definecolor{BlueGreen}{RGB}{49,152,255}
\definecolor{Violet}{RGB}{120,80,120}
\definecolor{Blue}{RGB}{0,0,255}
\definecolor{Yellow}{RGB}{0,255,51}
\definecolor{ElectricGreen}{RGB}{0, 255, 0}
\definecolor{MediumPersianBlue}{RGB}{0, 103, 165}

\usepackage[unicode, colorlinks, urlcolor=BlueGreen, linkcolor=Blue, citecolor=ElectricGreen]{hyperref}

\usepackage[left=2cm,right=2cm,top=2cm,bottom=2cm,bindingoffset=0cm]{geometry}
\numberwithin{equation}{section}

\makeatletter
\newcommand{\sgn}{ \operatorname{sgn} }

\makeatother

\usepackage{authblk}

\title{\bf Higher loop corrections to the current of created pairs in the lengthy electric pulse}
\author[1,2]{E.~T.~Akhmedov\thanks{\href{mailto:akhmedov@itep.ru}{akhmedov@itep.ru}}}
\author[1]{P.~S.~Zavgorodny\thanks{\href{mailto:zavgorodnij.ps@phystech.edu}{zavgorodnij.ps@phystech.edu}}}

\affil[1]{\textcolor{black}{Institutskii per, 9, Moscow Institute of Physics and Technology, 141700, Dolgoprudny, Russia}}
\affil[2]{\textcolor{black}{NRC ``Kurchatov Institute'', 123182, Moscow, Russia}}
\date{\today}

\begin{document}
\maketitle
\begin{abstract}
 The Schwinger's current appearing due to the breakdown of the vacuum in the strong constant electric pulse background is well known and calculated in many places. It appears due to the amplification of the zero-point fluctuations. However, we find that in higher loop corrections there are hidden redistributions of the level populations of charged particles and photons. These redistributions lead to the corrections to the current, which become large for the lengthy enough pulse and change the Schwinger's expression for the current.
\end{abstract}

\section{Introduction}\label{Introduction}

In his seminal paper \cite{Schwinger:1951nm} Schwinger has calculated particle creation rate in QED with constant electric field background $w \sim E^2e^2\, \exp\left[-\pi m^2/eE\right]$. In \cite{Grib:1976zw, Grib:1980aih, Zeldovich:1971mw, Nikishov, Nikishov:1969tt} generalizations of the Schwinger's calculation to other types of background electric fields were performed. In the spatially constant electric pulse background it is not hard to estimate the current  appearing due to the breakdown of the vacuum: the current along the field is $j \sim e \, w \, T$, where $T$ is the duration of the background pulse. This expression for the current was obtained in many places \cite{Gavrilov:1996pz,Kluger:1992gb,Gavrilov:2005dn,Gavrilov:2007hq,Gavrilov:2007ij,Gavrilov:2008fv,Gavrilov:2012jk,Anderson:2013zia,Anderson:2013ila} (see also \cite{Krotov:2010ma, Akhmedov:2014hfa, Akhmedov:2014doa, Akhmedov:2020dgc}).

If the duration of the background field continues for long enough, one can expect at least that created pairs, being accelerated, will radiate photons. Then, radiated photons can decay into pairs of the charged particles. In the background field such processes are not forbidden. The natural question is how to estimate effects of such processes on the current during the breakdown of the vacuum? Will the new processes contribute substantially to the current and drastically change the Schwinger's expression for the current? Or, how long should the background field act to make these effects feasible?

These questions being only of academic interest at present achievable values of strong electric fields can still shed some light for other types of strong background fields: e.g. gravitational ones, which very probably have been present in the very early Universe (see e.g. \cite{Krotov:2010ma}, \cite{Akhmedov:2013vka}, \cite{Akhmedov:2015xwa}, \cite{Akhmedov:2017ooy}, \cite{Akhmedov:2019cfd}, \cite{Akhmedov:2021rhq},\cite{Akhmedov:2022whm}, \cite{Akhmedov:2024lce} the related discussion). In this context our main interest is to find all these effects from first principles of quantum field theory rather than just present their estimates on general physical grounds, which are not well established at least for strong background fields, e.g. for $ m^2/e \leq E$.

The formulated above questions have been addressed in constant eternal electric field in \cite{Akhmedov:2014hfa, Akhmedov:2014doa}. The result was that the phenomena under consideration are hidden within infrared loop effects (appearing in the strong background field) and do indeed change the Schwinger's expression for the current, which follows from the tree-level propagators. Namely, the Schwinger current just follows from the amplification of the zero-point fluctuations. Meanwhile, in the loops one at least observes the redistribution of the level populations of the higher levels. That redistribution leads to a change of the current for the long enough time of the observation.

However, the constant eternal electric field background is not a physically realizable situation in a laboratory and is quite pathological. In particular, the tree-level current in the constant eternal electric field background is strictly vanishing for a reasonable (Hadamard, time-translation and time-reversal invariant) initial state. But in the loops the current is not zero and is infrared divergent, if one shifts the initial Cauchy surface to past infinity. The point is that even after the introduction of the UV cutoff loop corrections are infinite in the background fields, if the calculation is performed in the entire space-time. To cutoff the divergence one has to introduce the initial Cauchy surface, which violates the time-translation and time-reversal symmetries of the constant eternal electric field and of the initial state. That generates the non-zero growing with time current in the loops. The latter observation just hints that the answer on the formulated above questions depends on the initial conditions, which begs for the consideration of the pulse background rather than of the eternal field. 

In our previous paper \cite{Akhmedov:2023zfy} we have performed similar calculations in the electric pulse background. We have shown that one-loop corrections to the current, unlike the case of eternal field, bring nothing more than just the UV renormalization of the tree-level expression. (In our present paper we show that this happens just due to a tiny cancellation, which does not occur in higher loops.) However, corrections to the photon propagator contain a growth with the pulse duration and may affect higher loop corrections to the current. For this reason, in this paper, we consider higher loop corrections to the electric current and check their effect. We observe that higher loops grow faster than one would expect and do drastically change the current. (Exactly because of the absence of the aforementioned cancellation.)

For the integrity let us stress that large loop corrections in external fields are also often discussed in the context of the Ritus-Narozhny conjecture. One can find the discussion of the
increase of loop contributions with field strength in constant crossed fields ($\mathbf{E}\perp\mathbf{H},\; E=H$) \cite{Ritus1970,Narozhnyi:1980dc, Fedotov:2016afw} and constant magnetic field \cite{Karbstein:2019wmj}. However, these effects are not directly related to the main subject of our paper, because we are interested in the time evolution of the initial state of the theory and in the growth of the loop corrections with the duration of the pulse, rather than just a breakdown of perturbation theory and a growth of the loop corrections with the field strength.

\section{Set up of the problem}
\subsection{Action and modes}

In this paper we consider scalar electrodynamics with an external source $ j_{\mu}^{\text{cl}} $,
\begin{equation}\label{SQED_action_initial}
S[\phi,\phi^\dagger; A^\mu]=\int d^4x\left[
|\partial_\mu\phi+ieA_\mu\phi|^2-m^2|\phi|^2
-\dfrac{1}{4}F_{\mu\nu}F^{\mu\nu}-j_{\mu}^{\text{cl}}A^\mu \right],
\end{equation}
and divide the vector potential $ A^\mu $ into the classical and quantum parts, $ A^\mu = A_{\text{cl}}^\mu + a^\mu $, where $A_{\text{cl}}^\mu$ solves the classical equations of motion with the source. The source is such that the classical field is the lengthy, $eET^2\gg 1$, homogeneous electric pulse background,
\begin{equation}\label{background_field}
A^{\text{cl}}_\mu=(0; A_1(t); 0; 0),\quad A_1(t)=ET\tanh\dfrac{t}{T}.
\end{equation}
For the gauge field $ a_\mu $ we choose the Feynman gauge and quantize it in the standard way with $\widehat{\alpha}_{\mathbf{q}\mu}^\dagger$, $\widehat{\alpha}_{\mathbf{q}'\nu}$ being the creation and annihilation operators, correspondingly. The external electric field does not affect the free field quantization of photons. For the scalar field we have the following decomposition:
\begin{equation}\label{phi_decomposition}
\widehat{\phi}(t,\mathbf{x})=\int\dfrac{d^3\mathbf{p}}{(2\pi)^3}\left[
\widehat{a}_{\mathbf{p}}e^{i\mathbf{px}}f_{\mathbf{p}}(t)+
\widehat{b}^{\dagger}_{\mathbf{p}}e^{-i\mathbf{px}}f^{*}_{-\mathbf{p}}(t)
\right],
\end{equation}
where the temporal part of the mode function $f_{\mathbf{p}}(t)$ solves the equation
\begin{equation}\label{modes_with_background}
\left( \partial_t^2+\Big[\mathbf{p}+e\mathbf{A}(t)\Big]^2+m^2\right)f_{\mathbf{p}}(t)=0.
\end{equation}
We require the harmonic function $f_{\mathbf{p}}(t)$ to be the in-mode, i.e. the single wave at the past infinity:
\begin{equation}\label{asympt_past}
f^{\text{in}}_\mathbf{p}(t/T\to-\infty)\simeq\dfrac{e^{-i\omega_{-}t}}{\sqrt{2\omega_{-}}},\quad
\omega_\pm(\mathbf{p})=\sqrt{(\mathbf{p}+\mathbf{A}(\pm\infty))^2+m^2}.
\end{equation}
The exact form of the modes through the hypergeometric functions is not necessary for us below. Meanwhile, the asymptotic form of the modes $ f_\mathbf{p}(t) $ in the future infinity, $ t/T\to\infty $, is as follows:
\begin{equation}\label{MathcalApm1}
f_{\mathbf{p}}^{\mathrm{in}}(t / T \rightarrow+\infty)=\mathcal{A}_{+}(\mathbf{p}) e^{i \omega_{+} t}+\mathcal{A}_{-}(\mathbf{p}) e^{-i \omega_{+} t},
\end{equation}
where
\begin{equation}\label{MathcalApm}
|\mathcal{A}_{-}(\mathbf{p})|^2=\dfrac{1}{2\omega_+(\mathbf{p})}+
|\mathcal{A}_{+}(\mathbf{p})|^2,\quad 
|\mathcal{A}_{+}(\mathbf{p})|^2=\dfrac{1}{2\omega_+(\mathbf{p})}\cdot
\exp \left[-\frac{\pi\left(\mathbf{p}_{\perp}^2+m^2\right)}{e E}\right]
\end{equation}
and $ \mathbf{p}_\perp=(p_2,p_3) $.

For the future reference let us stress that for $|t|<T $ the mode function $ f_\mathbf{p}(t) $ can be approximated up to the order of $ \mathcal{O}(1/T^2) $ by the parabolic cylinder function,
\begin{equation}\label{straight_modes}
f_{\mathbf{p}}(t) \simeq g_{\mathbf{p}_\perp}(p_1+eEt) = c(\mathbf{p}_\perp) D_{-\frac{1}{2}+i\frac{\mathbf{p}^2_\perp+m^2}{2eE}}
\left[-\dfrac{(1-i)(p_1+eEt)}{\sqrt{eE}} \right],
\end{equation}
where 
\begin{equation}\label{cylinder_coefs}
c(\mathbf{p}_\perp)=\dfrac{2^{ieET^2}}{2^{\xi_0/2}[eE]^{1/4}}
\left[eET^2\right]^{-i\frac{\mathbf{p}^2_\perp+m^2}{4eE}}\cdot
e^{-i\pi/8}\cdot\exp\left[-\dfrac{\pi(\mathbf{p}^2_\perp+m^2)}{8eE} \right].
\end{equation}
The initial state that we consider in the calculations below is the Fock space in-ground state: $\widehat{a}_{\mathbf{p}} \, |in \rangle = 0$.

In the present paper we are interested in the calculation of the level populations, anomalous quantum expectation values (for both photons and charged complex scalars) and electric currents {\it during the electric pulse duration}. The point is that when the pulse is switched off we are dealing with the standard scalar QED, i.e. without any background field, but with an unusual initial state (right after the pulse is switched off) due to the presence of created scalars, photons and the anomalous expectation values for the corresponding fields. Such a situation is similar to the one considered in \cite{Akhmedov:2021vfs}, but in a bit different theory. In any case, on general grounds one can expect that after the switching off of the background electric field there will be a thermalization process, at least for small enough anomalous expectation values. But that is a different physical situation and, thus, a different problem, which of course can have its own specifics and academic interest. But we are interested in the kinetics during the electric pulse duration. The fact that we consider particle creation process during the pulse is the reason why in the expressions below we frequently use the mode functions in the approximate form $g_{\mathbf{p}_\perp}(p_1+eEt)$ rather that $f_{\mathbf{p}}(t)$.

\subsection{Tree-level current and one-loop corrections}

Tree-level expression for the current in the long pulse, $eET^2\gg 1$, in the future infinity, $ t/T\gg 1 $, has been calculated in many places both for the scalar and standard (spinor) QED \cite{Grib:1980aih, Gavrilov:1996pz, Gavrilov:2007hq, Gavrilov:2008fv, Akhmedov:2020dgc} and has the following form:
\begin{equation}\label{tree_current_renorm_finally}
j_1(t) \simeq\dfrac{E^2e^3T}{2\pi^3}\cdot\exp\left[-\dfrac{\pi m^2}{eE}\right].
\end{equation}
This expression has a clear physical explanation: during the pulse the pairs of charged particles are created with the Schwinger's probability rate per unit four-volume $w \sim E^2e^2\, \exp\left[-\pi m^2/eE\right]$. Hence, by the end of the pulse duration one obtains the current density of the form $j_1 \sim e \, w \, T$. Our goal here is to check if loop corrections modify this expression substantially for long enough pulse.

\begin{figure}[h]
\center{\includegraphics[width=0.6\linewidth]{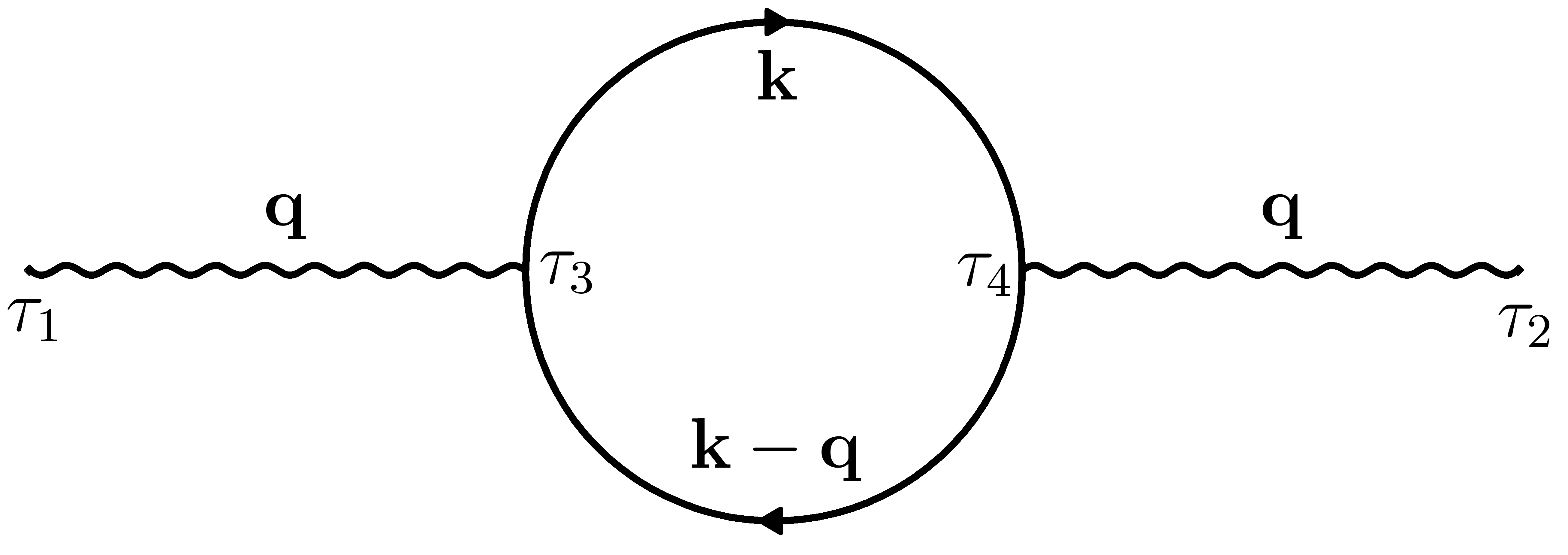}}
\caption{One-loop correction to the photon's propagator. In the Schwinger-Keldysh technique one has a sum of diagrams of such a form.}
\label{image_dia_photon_0}
\end{figure}

In \cite{Akhmedov:2023zfy} it was shown that
one-loop correction to the tree-level current being suppressed by the extra power of the fine structure constant, $ e^2 $, in comparison with the tree-level current, does not grow faster than (\ref{tree_current_renorm_finally}) with $T$. Hence, even for a lengthy pulse the tree--level current is not drastically modified by the first quantum loop correction.

However, in \cite{Akhmedov:2023zfy} it was also shown that the one-loop correction to the photon's Keldysh propagator, shown in Fig. \ref{image_dia_photon_0}, does grow\footnote{For the constant background electric field such a secular growth (actually secular divergence) was found in \cite{Akhmedov:2014hfa,Akhmedov:2014doa}.} with $ T $. Namely, the correction to the photon's Keldysh propagator in the limit $ |\tau_1-\tau_2|\ll|\tau_1+\tau_2|/2 \equiv |\mathcal{T}| < T$ has the following form:
\begin{equation}\label{K_photon}
\Delta G^K_{\mu\nu}(\mathbf{q};\tau_1,\tau_2)\simeq n_{\mu\nu}(\mathbf{q},\mathcal{T})\left[ 
\dfrac{e^{-i|\mathbf{q}|(\tau_1-\tau_2)}}{2|\mathbf{q}|}+\operatorname{h.c.}
\right],
\end{equation}
where $ n_{\mu\nu}(\mathbf{q}) \, \delta \left(\mathbf{q} - \mathbf{q}'\right) = \left\langle \widehat{\alpha}_{\mathbf{q}\mu}^\dagger
\widehat{\alpha}_{\mathbf{q}'\nu}\right\rangle $ and $\widehat{\alpha}_{\mathbf{q}\mu}^\dagger$, $\widehat{\alpha}_{\mathbf{q}'\nu}$ are photon's creation and annihilation operators, correspondingly. In the leading order in powers of large $T$ and $|\mathcal{T}| < T$ we have
\begin{equation}\label{n_munu-growth}
\begin{split}
n_{\mu\nu}(\mathbf{q},\mathcal{T}) \approx -2e^2(\mathcal{T}+T)\int\limits_{-\infty}^{+\infty}d\tau
\dfrac{e^{-2i|\mathbf{q}|\tau}}{2|\mathbf{q}|}\int\dfrac{d^3\mathbf{k}}{(2\pi)^3}\times \\
\times \mathbb{D}_{\mu\nu}\left[
g_{\mathbf{k}_\perp}(k_1+eE\tau)g^*_{\mathbf{k}_\perp}(k_1-eE\tau);
g^*_{\mathbf{k}_\perp-\mathbf{q}_\perp}(k_1-q_1-eE\tau)g_{\mathbf{k}_\perp-\mathbf{q}_\perp}(k_1-q_1+eE\tau)
\right],
\end{split}
\end{equation}
where the function $g_{\mathbf{k}_\perp}(k_1+eE\tau)$ is defined in eq. (\ref{straight_modes}) and the operator $ \mathbb{D}_{\mu\nu} $ acts as
\begin{equation}\label{mathD}
\begin{aligned}
\mathbb{D}_{\mu\nu}\left[
g_{\mathbf{k}_\perp}(k_1+eE\tau)g^*_{\mathbf{k}_\perp}(k_1-eE\tau);
g^*_{\mathbf{k}_\perp-\mathbf{q}_\perp}(k_1-q_1-eE\tau)g_{\mathbf{k}_\perp-\mathbf{q}_\perp}(k_1-q_1+eE\tau)
\right] = \\
=\left[D_\mu g_{\mathbf{k}_{\perp}}\left(k_1+e E \tau\right) g_{\mathbf{k}_{\perp}-\mathbf{q}_{\perp}}\left(k_1-q_1+e E \tau\right)-
g_{\mathbf{k}_{\perp}}\left(k_1+e E \tau\right) D_\mu^\dagger g_{\mathbf{k}_{\perp}-\mathbf{q}_{\perp}}\left(k_1-q_1+e E \tau\right)\right]\times \\
\times
\left[g_{\mathbf{k}_{\perp}}^*\left(k_1-e E \tau\right) D_\nu g_{\mathbf{k}_{\perp}-\mathbf{q}_{\perp}}^*\left(k_1-q_1-e E \tau\right)-
D_\nu^{\dagger}g_{\mathbf{k}_{\perp}}^*\left(k_1-e E \tau\right)g_{\mathbf{k}_{\perp}-\mathbf{q}_{\perp}}^*\left(k_1-q_1-e E \tau\right)\right].
\end{aligned}
\end{equation}
The expression (\ref{n_munu-growth}) explicitly grows with $ \mathcal{T} $ and $ T $. Such a growth of the photon level-population is due to the fact that photons are created together with scalar particles from the initial state. The process in question is present in the interacting theory and is due to the presence of the vertex $a_\mu |\phi|^2$. Meanwhile the anomalous quantum averages for the photon field, $\langle \alpha \, \alpha\rangle$ and $\langle \alpha^+ \, \alpha^+\rangle$, do not grow with time, which is in agreement with the expectation that the initial Fock space ground state for the photon field remains to be the ground state of the Hamiltonian of the free photon theory.

\begin{figure}[h]
\center{\includegraphics[width=0.6\linewidth]{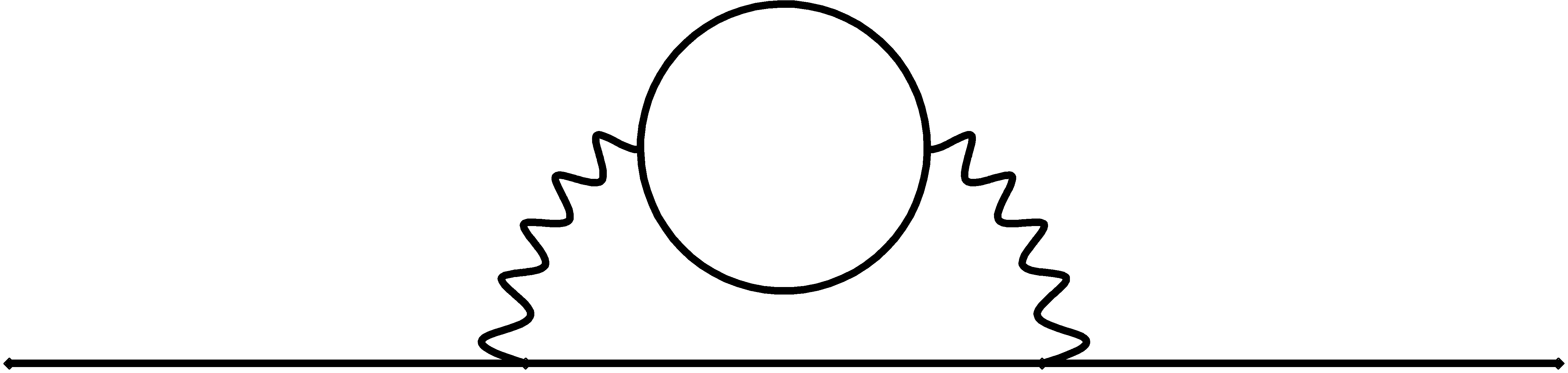}}
\vspace{+1ex}
\caption{This two-loop diagram contains additional power of growth with $ T $. In the Schwinger-Keldysh technique one has a sum of diagrams of such a form.}
\label{image_dia_photon}
\end{figure}

In any case, due to this growth of the photon's two-point function we expect that in the two-loop order the correction to the current of the created pairs (from the diagram shown on the Fig. \ref{image_dia_photon}), although being suppressed by the extra power of the fine structure constant, will bring an additional power of growth with $ T $ (higher than linear growth, which is present in the tree-level contribution to the current). The goal of this paper is to check if this expectation is correct, check the contributions of other two-loop diagrams and discuss physical consequences of these observations. 

\section{Two-loop corrections to the current}

In this section we estimate two-loop corrections to the two-point correlation functions. We begin with the diagrams of the form shown on the Fig. \ref{image_dia_photon}, since, as we mentioned above, such diagrams potentially can bring growing faster with $T$ contributions to the current in comparison with the tree-level expression (\ref{tree_current_renorm_finally}), which grows linearly.

\subsection{Leading two-loop diagram}

It is convenient to rewrite the contribution coming from the diagrams shown on the Fig. \ref{image_dia_photon} in terms of the advanced, retarded and Keldysh propagators,
\begin{equation}
G^R=G^{--}-G^{-+},\quad
G^A=G^{--}-G^{+-},\quad
G^K=\dfrac{1}{2}\Big[ G^{-+}+G^{+-} \Big].
\end{equation}
In this form two-loop correction to the current has the following structure:
\begin{equation}\label{corrcurr}
\begin{aligned}
\Delta j_1(t) = 2e\int\dfrac{d^3\mathbf{p}}{(2\pi)^3}(p_1+eET)\Delta D^K(\mathbf{p};t,t) &= 2e\int\dfrac{d^3\mathbf{p}}{(2\pi)^3}(p_1+eET)\int d\tau_1 d\tau_2
\int \dfrac{d^3\mathbf{q}}{(2\pi)^3} \times \\
&\times\mathbb{D}^{\mu\nu}\Big{[}
D^{R}(\mathbf{p};t,\tau_1)
\Sigma^{K}(\mathbf{p},\mathbf{q};\tau_1,\tau_2)
D^{A}(\mathbf{p};\tau_2,t)+ \\
&+
D^{K}(\mathbf{p};t,\tau_1)
\Sigma^{A}(\mathbf{p},\mathbf{q};\tau_1,\tau_2)
D^{A}(\mathbf{p};\tau_2,t)+ \\
&+
D^{R}(\mathbf{p};t,\tau_1)
\Sigma^{R}(\mathbf{p},\mathbf{q};\tau_1,\tau_2)
D^{K}(\mathbf{p};\tau_2,t)\Big{]},
\end{aligned}
\end{equation}
where $\mathbb{D}^{\mu\nu}$ acts on mode functions of the scalar field as in (\ref{mathD}), and $ \Sigma $ is the self-energy, which in the Schwinger-Keldysh technique is the matrix of the form:
\begin{equation}
\begin{pmatrix}
2\Sigma^K & \Sigma^R \\
\Sigma^A & 0
\end{pmatrix}=-
\begin{pmatrix}
2D^K\Delta G^K+\frac{1}{2}\left[D^A\Delta G^A+D^R\Delta G^R\right] & D^K\Delta G^R+D^R\Delta G^K \\
D^K\Delta G^A+D^A\Delta G^K & 0
\end{pmatrix},
\end{equation} 
where $ D^{K,R,A}=D^{K,R,A}(\mathbf{p}-\mathbf{q};\tau_1,\tau_2) $ is the tree-level Keldysh (retarded, advanced) propagators of the scalar field, while $ \Delta G^{K(R,A)}= \Delta G^{K(R,A)}_{\mu\nu}(\mathbf{ q};\tau_1,\tau_2) $ is the one-loop corrected photon's Keldysh (retarded, advanced) propagators. To obtain this correction to the current (\ref{corrcurr}) we use the contribution to the Keldysh propagator coming from the diagram shown on the Fig. \ref{image_dia_photon} at the coincident space-time points, as is expressed in the first row of (\ref{corrcurr}).

Only those terms that contain $ \Delta G^K_{\mu\nu}(\mathbf{ q};\tau_1,\tau_2) $ contribute the leading corrections at large $T$, since it is only the Keldysh propagator that contains the extra power of $T$, as is shown in \cite{Akhmedov:2023zfy}, and we have discussed above. Our goal is to single out the leading contribution for the large $T$.

In all, the leading correction to the current in the large $T$ limit can be written as:
\begin{equation}\label{Current-resummation}
\begin{split}
\Delta j_1 & \approx -2e\int\dfrac{d^3\mathbf{p}}{(2\pi)^3}(p_1+eET)\int d\tau_1 d\tau_2
\int \dfrac{d^3\mathbf{q}}{(2\pi)^3} \, \Delta G^K_{\mu\nu}(\mathbf{q};\tau_1,\tau_2)\times \\
&\times\mathbb{D}^{\mu\nu}\Big{[}
D^{R}(\mathbf{p};t,\tau_1)
D^{K}(\mathbf{p}-\mathbf{q};\tau_1,\tau_2)
D^{A}(\mathbf{p};\tau_2,t)+ \\
&+
D^{K}(\mathbf{p};t,\tau_1)
D^{A}(\mathbf{p}-\mathbf{q};\tau_1,\tau_2)
D^{A}(\mathbf{p};\tau_2,t)+ \\
&+
D^{R}(\mathbf{p};t,\tau_1)
D^{R}(\mathbf{p}-\mathbf{q};\tau_1,\tau_2)
D^{K}(\mathbf{p};\tau_2,t)\Big{]}.
\end{split}
\end{equation}
After the substitution of the expression for the tree-level propagators $ D^{R,K,A} $ via the mode functions, we can rewrite the leading correction to the current as
\begin{equation}\label{exact_current_finally}
\Delta j_1(t) = 2e\int\dfrac{d^3\mathbf{p}}{(2\pi)^3}\left(p_1+eET\right)\left[
\left|f_\mathbf{p}(t)\right|^2 n_\mathbf{p}(t)+
\left(f^*_\mathbf{p}(t)\right)^2 \kappa_\mathbf{p}(t)+
\left(f_\mathbf{p}(t)\right)^2 \kappa^*_\mathbf{p}(t)
\right],
\end{equation}
where
\begin{equation}\label{n-p(t)-2}
\begin{split}
n_\mathbf{p}(t) &\simeq -2e^2\int\limits_{t_0}^{t}d\tau_1\int\limits_{t_0}^{t}d\tau_2\int\dfrac{d^3\mathbf{q}}{(2\pi)^3}
\Delta G^{K}_{\mu\nu}(\mathbf{q};\tau_1,\tau_2) \times \\
&\times \mathbb{D}^{\mu\nu}\bigg[
f^*_\mathbf{p}(\tau_1);\;
f^*_{\mathbf{p}-\mathbf{q}}(\tau_1)
f_{\mathbf{p}-\mathbf{q}}(\tau_2);\;
f_\mathbf{p}(\tau_2)
\bigg]
\end{split}
\end{equation}
and
\begin{equation}\label{kappa-p(t)-2}
\begin{split}
\kappa_\mathbf{p}(t) &\simeq 2e^2\int\limits_{t_0}^{t}d\tau_1\int\limits_{t_0}^{\tau_1}d\tau_2\int\dfrac{d^3\mathbf{q}}{(2\pi)^3}
\Delta G^{K}_{\mu\nu}(\mathbf{q};\tau_1,\tau_2) \times \\
&\times \mathbb{D}^{\mu\nu}\bigg[
f_\mathbf{p}(\tau_1);\;
f^*_{\mathbf{p}-\mathbf{q}}(\tau_1)
f_{\mathbf{p}-\mathbf{q}}(\tau_2);\;
f_\mathbf{p}(\tau_2)
\bigg].
\end{split}
\end{equation}
These expressions can be used both during the pulse duration and after its switching off. That is the reason why in them we denote the mode functions as $f_\mathbf{p}(t)$ rather than $g_{\mathbf{p}_\perp}(p_1+eEt)$. The physical meaning of the obtained expression for the correction to the current is clear: while the tree-level current (\ref{tree_current_renorm_finally}) appears due to the amplification of the zero-point fluctuations, the correction to the current (\ref{exact_current_finally}) appears due to the excitation of the higher levels and due to the generation of the anomalous quantum expectation values. 

To estimate the obtained expression in (\ref{n-p(t)-2}) and (\ref{kappa-p(t)-2}) we make the following change of integration variables,
\begin{equation}\label{varchange}
\tau=\frac{\tau_1-\tau_2}{2}, \quad \mathcal{T}=\frac{\tau_1+\tau_2}{2},
\end{equation}
and recall, that the propagator $ \Delta G^{K}_{\mu\nu}(\mathbf{q};\tau_1,\tau_2) $ contains growth with $ T $ only for $ |\tau|\ll|\mathcal{T}| $, where it has the form (\ref{K_photon}). Due to the oscillating functions in (\ref{K_photon}) the integral over $ \mathbf{q} $ for large $ \tau $ is saturated within a short region. Hence, we can extend the integration over $ \tau $ onto the entire real axis. 

As we explained above, we are interested only in the time region inside the pulse duration. Then, we restrict the integration over $ \mathcal{T} $ to the region $ |\mathcal{T}|<T $. This allows us to use the parabolic cylinder function approximation (\ref{straight_modes}) for the modes and apply the expression (\ref{n_munu-growth}) for $ n_{\mu\nu}(\mathbf{q},\mathcal{T}) $. After the change of variables,
\begin{gather}\label{dimensionless_vars}
\chi=\dfrac{p_1+eET}{\sqrt{eE}},\quad
\chi_i=\dfrac{p_1+eE\tau_i}{\sqrt{eE}},\quad
\mathbf{Q}=\dfrac{\mathbf{q}}{\sqrt{eE}},
\nonumber
\\
X=\dfrac{\chi_1+\chi_2}{2},\quad \widetilde{\chi}=\chi_1-\chi_2,
\end{gather}
we obtain the following expressions:
\begin{equation}\label{2-loop-n_p(T)}
\begin{split}
n_\mathbf{p}(T)\simeq \mathcal{N}_{\mathbf{p}_\perp}(\chi) &= -2e^2
\int\limits_{\chi-2\sqrt{eE}T}^{\chi}dX
\int\limits_{-\infty}^{+\infty}d\widetilde{\chi}
\int\dfrac{d^3\mathbf{Q}}{(2\pi)^3}
\Delta G^{K}_{\mu\nu}(\mathbf{Q};X,\widetilde{\chi})\times \\
&\times \mathbb{D}^{\mu\nu}\bigg[
g^*_{\mathbf{p}_\perp}(X+\widetilde{\chi}/2)
g^*_{\mathbf{p}_\perp-\mathbf{q}_{\perp}}(X+\widetilde{\chi}/2-Q_1)\times \\
&\times
g_{\mathbf{p}_\perp-\mathbf{q}_{\perp}}(X-\widetilde{\chi}/2-Q_1)
g_{\mathbf{p}_\perp}(X-\widetilde{\chi}/2)\bigg],
\end{split}
\end{equation}
and
\begin{equation}\label{2-loop-kappa_p(T)}
\begin{split}
\kappa_\mathbf{p}(T)\simeq \mathcal{K}_{\mathbf{p}_\perp}(\chi) &= 2e^2
\int\limits_{\chi-2\sqrt{eE}T}^{\chi}dX
\int\limits_{0}^{+\infty}d\widetilde{\chi}
\int\dfrac{d^3\mathbf{Q}}{(2\pi)^3}
\Delta G^{K}_{\mu\nu}(\mathbf{Q};X,\widetilde{\chi})\times \\
&\times \mathbb{D}^{\mu\nu}\bigg[
g_{\mathbf{p}_\perp}(X+\widetilde{\chi}/2)
g^*_{\mathbf{p}_\perp-\mathbf{q}_{\perp}}(X+\widetilde{\chi}/2-Q_1)\times \\
&\times
g_{\mathbf{p}_\perp-\mathbf{q}_{\perp}}(X-\widetilde{\chi}/2-Q_1)
g_{\mathbf{p}_\perp}(X-\widetilde{\chi}/2)\bigg].
\end{split}
\end{equation}
We want to extend the integration over $ X $ to the entire real axis using the asymptotic form of the parabolic cylinder function with the large argument,
\begin{equation}\label{Cylinder_2}
g_{\mathbf{p}_\perp}(X)\simeq
\alpha_\pm(\mathbf{p}_\perp)\dfrac{X^{i\frac{m^2+\mathbf{p}^2_\perp}{2eE}}}{X^{1/2}}
e^{\frac{iX^2}{2}}+
\beta_\pm(\mathbf{p}_\perp)\dfrac{X^{-i\frac{m^2+\mathbf{p}^2_\perp}{2eE}}}{X^{1/2}}
e^{\frac{-iX^2}{2}},
\end{equation} 
where the sign $\pm$ corresponds to the large positive and negative $X$ values, correspondingly, and
\begin{equation}
\begin{aligned}
& \left|\alpha_{+}\left(\mathbf{p}_{\perp}\right)\right|^2=\frac{1}{2 \sqrt{e E}} \exp \left[-\frac{\pi\left(\mathbf{p}_{\perp}^2+m^2\right)}{e E}\right], \quad\left|\alpha_{-}\left(\mathbf{p}_{\perp}\right)\right|^2=\frac{1}{2 \sqrt{e E}} \\
& \left|\beta_{+}\left(\mathbf{p}_{\perp}\right)\right|^2=\frac{1}{2 \sqrt{e E}}+\left|\alpha_{+}\left(\mathbf{p}_{\perp}\right)\right|^2, \quad\left|\beta_{-}\left(\mathbf{p}_{\perp}\right)\right|^2=0.
\end{aligned}
\end{equation}
We can extend the lower limit of integration to minus infinity, since there is only one exponent in (\ref{Cylinder_2}) for $ X<0 $ and the integrands in (\ref{2-loop-n_p(T)}), (\ref{2-loop-kappa_p(T)}) are fast oscillating functions in the asymptotic past. 

However, there are several obstacles to extend the upper limit to plus infinity. First, the photon propagator contains additional growth with $ X $. Second, it is not clear whether $ \mathbb{D}^{\mu\nu} $ contains a growth with $X$ after the convolution with $ \Delta G_{\mu\nu} $ or not. Therefore, we examine in more detail the operator $ \mathbb{D}^{\mu\nu} $ in the asymptotic region $ 1\ll X<\chi $:
\begin{equation}
\mathbb{D}^{\mu \nu}=\left[\left[D^\mu\right]_1^{\dagger}-D_1^\mu\right] \cdot\left[\left[D^\nu\right]_2^{\dagger}-D_2^\nu\right],
\end{equation}  
where
\begin{equation}\label{D-D-11}
\left[D^1\right]_1^{\dagger}-D_1^1=i(2X-Q_1+\widetilde{\chi})\sqrt{eE};\quad
\left[D^1\right]_2^{\dagger}-D_2^1=i(2X-Q_1-\widetilde{\chi})\sqrt{eE},
\end{equation}
and
\begin{equation}\label{D-D-0}
\left[D^0\right]_1^{\dagger}-D_1^0=\pm i(2X-Q_1+\widetilde{\chi})\sqrt{eE};\quad
\left[D^0\right]_2^{\dagger}-D_2^0=\pm i(2X-Q_1-\widetilde{\chi})\sqrt{eE}.
\end{equation}
From (\ref{D-D-11}) and (\ref{D-D-0}) it is clear that the trace $ \mathbb{D}=\mathbb{D}^\mu_\mu $, which is contained in the one-loop correction to the current, is independent\footnote{ That actually was the reason why the first loop correction to the current did not grow faster than the first power of $T$.} of $ X $. But in (\ref{2-loop-n_p(T)}), the convolution is less trivial and does not reduce to the combination $ \mathbb{D}^0_0+\mathbb{D}^1_1 $, i.e. to the trace. Due to components of $ \mathbb{D}^{\mu\nu} $ with $ \mu,\nu=0,1 $, we have an additional growth of order $ X^2 $ in (\ref{2-loop-n_p(T)}). The argument of the operator $ \mathbb{D}^{\mu\nu} $ is of the order of $ 1/X^2 $, as follows from (\ref{Cylinder_2}). Thus, we find that the integral over $ X $ can be of the order of $T^2 $, which means that the correction to the electric current potentially contains even faster growth with $ T $, than we expected in the second loop order due to the linear growth of the photon's self-energy.

To check the power of the growth of the two-loop correction to the current with $T$ one has to estimate the integrals in (\ref{2-loop-n_p(T)}) and (\ref{2-loop-kappa_p(T)}) for the more explicit form of the mode functions. In our forthcoming publications we will do that separately in two limits: for the weak and strong background electric fields. In particular, in our next paper we will show that for the weak field background the growth with $T$ is actually faster than quadratic.

At this moment we see that the two loop correction being suppressed by the extra power of the fine structure constant is growing faster than $T^2$, which is faster than the linear growth of the tree-level contribution to the current (\ref{tree_current_renorm_finally}). This is the main conclusion of the present paper: for the lengthy enough background pulse the Schwinger's expression for the current of the created particles is drastically modified. Below we estimate other two-loop contributions to the current.

\subsection{Other two-loop corrections to the current}
We start with three two-loop diagrams, which also can potentially give an additional power of $ T $ to the current. They are shown on the Fig. \ref{image-two-loop}, \ref{image_dia_shifted_sunset} and \ref{image_dia_tadpole_double}.
Then we will continue with other types of two-loop diagrams which in principle can be present.

\subsubsection{Sunset diagram}
\begin{figure}[h]
\centering
\includegraphics[width=0.7\linewidth]{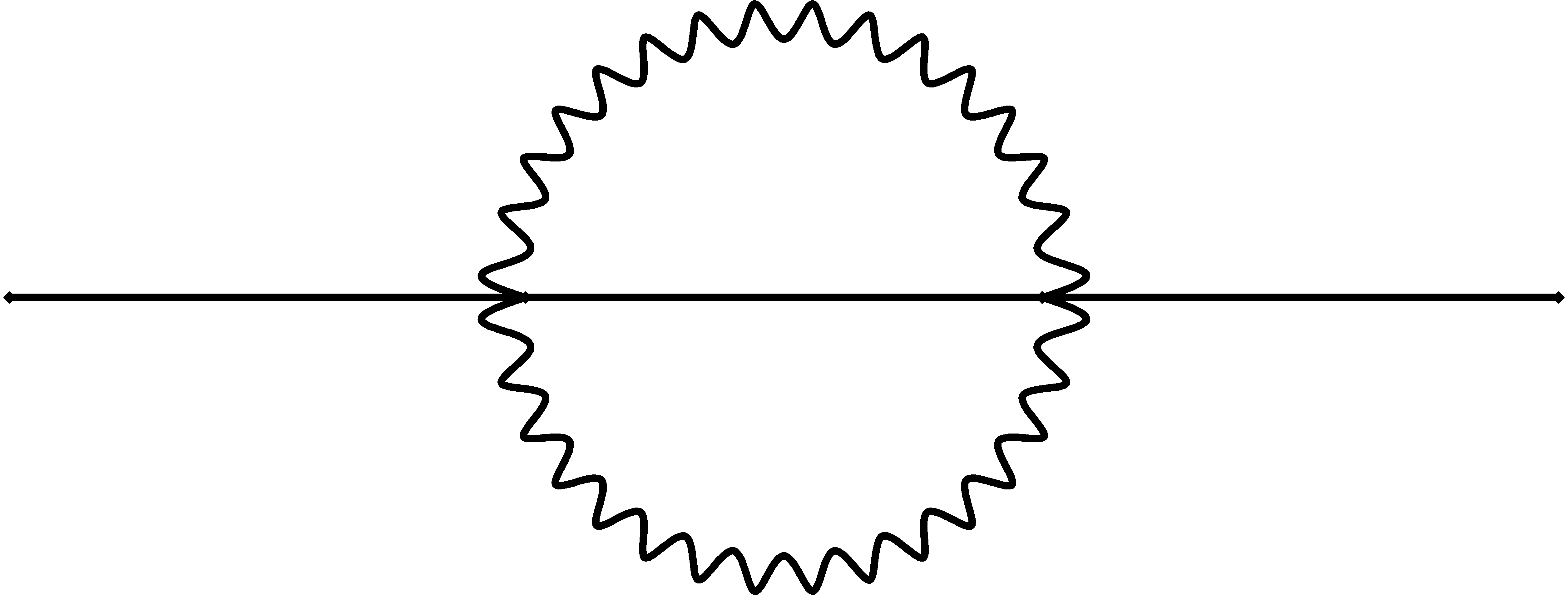}
\caption{Two-loop diagrams, which potentially can contain additional power of $ T $. In Schwinger-Keldysh technique one has a sum of such diagrams.}
\label{image-two-loop}
\end{figure}

Let us start with the consideration of the sunset diagram from the Fig. \ref{image-two-loop}. 
This correction to the scalar propagator in the momentum space can be expressed as follows:

\begin{equation}\label{loop_finally_dia}
\begin{split}
\Delta D^{ab}(\mathbf{p};t_1,t_2) &= 4\sum\limits_{c,d}\sgn(c)\sgn(d)e^4\int d\tau_1\int d\tau_2
\int\dfrac{d^3\mathbf{q}_1}{(2\pi)^3}
\int\dfrac{d^3\mathbf{q}_2}{(2\pi)^3}\times \\
&\times G^{cd}(\mathbf{q}_1,\tau_1,\tau_2)
G^{cd}(\mathbf{q}_2,\tau_1,\tau_2)\times \\
&\times 
D^{ac}(\mathbf{p},t_1,\tau_1)
D^{cd}(\mathbf{p}-\mathbf{q}_1-\mathbf{q}_2,\tau_1,\tau_2)
D^{db}(\mathbf{p},\tau_2,t_2),
\end{split}
\end{equation}
where $ a,b=\pm $ are the standard indexes in the Schwinger-Keldysh technique. Since the product of the two tree--level photon propagators has a similar time dependence as the single one, the procedure of the calculation of this two-loop diagram does not differ much from the one-loop calculation performed in \cite{Akhmedov:2023zfy}. The correction to the current again can be expressed in the standard form (\ref{exact_current_finally}), where
\begin{equation}\label{n-p(t)}
\begin{split}
n_\mathbf{p}(t) &\simeq -2e^4
\int\limits_{t_0}^{t}d\tau_1
\int\limits_{t_0}^{t}d\tau_2
\int\dfrac{d^3\mathbf{q}_1}{(2\pi)^3}
\int\dfrac{d^3\mathbf{q}_2}{(2\pi)^3}
\dfrac{e^{i(|\mathbf{q}|_1+|\mathbf{q}|_2)(\tau_1-\tau_2)}}
{|\mathbf{q}_1|\cdot|\mathbf{q}_2|}\times \\
&\times
f^*_\mathbf{p}(\tau_1)
f^*_{\mathbf{p}-\mathbf{q}_1-\mathbf{q}_2}(\tau_1)
f_{\mathbf{p}-\mathbf{q}_1-\mathbf{q}_2}(\tau_2)
f_\mathbf{p}(\tau_2),
\end{split}
\end{equation}
and
\begin{equation}\label{kappa-p(t)}
\begin{split}
\kappa_\mathbf{p}(t) &\simeq 2e^4
\int\limits_{t_0}^{t}d\tau_1
\int\limits_{t_0}^{\tau_1}d\tau_2
\int\dfrac{d^3\mathbf{q}_1}{(2\pi)^3}
\int\dfrac{d^3\mathbf{q}_2}{(2\pi)^3}
\dfrac{e^{i(|\mathbf{q}|_1+|\mathbf{q}|_2)(\tau_1-\tau_2)}}
{|\mathbf{q}_1|\cdot|\mathbf{q}_2|}\times \\
&\times
f_\mathbf{p}(\tau_1)
f^*_{\mathbf{p}-\mathbf{q}_1-\mathbf{q}_2}(\tau_1)
f_{\mathbf{p}-\mathbf{q}_1-\mathbf{q}_2}(\tau_2)
f_\mathbf{p}(\tau_2).
\end{split}
\end{equation}
After such a change of variables as (\ref{varchange})
we can extend the integration over $\tau$ in (\ref{n-p(t)}) onto the entire real axis, as we did above:

\begin{equation}\label{n-p(t)-1}
\begin{split}
n_\mathbf{p}(t) &\simeq -4e^4
\int\limits_{t_0}^{t}d\mathcal{T}
\int\limits_{-\infty}^{+\infty}d\tau
\int\dfrac{d^3\mathbf{q}_1}{(2\pi)^3}
\int\dfrac{d^3\mathbf{q}_2}{(2\pi)^3}
\dfrac{e^{2i(|\mathbf{q}|_1+|\mathbf{q}|_2)\tau}}
{|\mathbf{q}_1|\cdot|\mathbf{q}_2|}\times \\
&\times
f^*_\mathbf{p}(\tau_1)
f^*_{\mathbf{p}-\mathbf{q}_1-\mathbf{q}_2}(\tau_1)
f_{\mathbf{p}-\mathbf{q}_1-\mathbf{q}_2}(\tau_2)
f_\mathbf{p}(\tau_2),
\end{split}
\end{equation}
and similarly for the anomalous quantum average.

Let us stress at this moment that in contrast to the one-loop correction to the current, which was considered in \cite{Akhmedov:2023zfy}, in (\ref{n-p(t)-1}) there is a secular growth with $ t $ if $ t\gg T $. Please do not confuse the latter one with the growth with $T$, i.e. this time we discuss the growing with time change in the level population and anomalous average
after the switch off of the background field. This secular growth apparently is due to the fact that delta-functions establishing the energy conservation are no longer present. In fact, if the modes $f_{\bf p}(t)$ were single exponents, $e^{- i \, \omega(\mathbf{p}) \, t}/\sqrt{2\, \omega(\mathbf{p})}$, then in (\ref{n-p(t)-1}) from the integral
$$
\int\limits_{-\infty}^{+\infty}d\tau
e^{2i(|\mathbf{q}|_1+|\mathbf{q}|_2)\tau}
f^*_\mathbf{p}(\tau_1)
f^*_{\mathbf{p}-\mathbf{q}_1-\mathbf{q}_2}(\tau_1)
f_{\mathbf{p}-\mathbf{q}_1-\mathbf{q}_2}(\tau_2)
f_\mathbf{p}(\tau_2),
$$
one would obtain $\delta\Big(|\mathbf{q}_1|+|\mathbf{q}_2|+\omega(\mathbf{p})
+\omega(\mathbf{p}-\mathbf{q}_1-\mathbf{q}_2)\Big)$. The argument of the latter delta-function is never zero and, hence, $n_{\bf p}(t)$ is vanishing as it should be in the absence of the strong background field. The same is true for $\kappa_{\bf p}(t)$. But since in the presence of the strong background field the mode functions, $f_{\bf p}(t)$, are no longer pure exponents one obtains a smearing of the delta-function establishing the energy conservation: in fact, in the time dependent background under consideration there is no energy conservation. E.g. in the limit $t = (t_1 + t_2)/2 \gg |t_1-t_2|$ from (\ref{n-p(t)-1}) we approximately obtain the following contributions to the level population:

\begin{equation}\label{n-p(t)-infinity}
\begin{split}
n_\mathbf{p}(t) &\simeq -2e^4t
\int\dfrac{d^3\mathbf{q}_1d^3\mathbf{q}_2}
{(2\pi)^5|\mathbf{q}_1||\mathbf{q}_2|}
\left|\mathcal{A}_+(\mathbf{p})
\mathcal{A}_+(\mathbf{p}-\mathbf{q}_1-\mathbf{q}_2)\right|^2\times \\
&\times
\delta\Big(|\mathbf{q}_1|+|\mathbf{q}_2|-\omega_+(\mathbf{p})
-\omega_+(\mathbf{p}-\mathbf{q}_1-\mathbf{q}_2)\Big).
\end{split}
\end{equation}
To fetch this expression we used in (\ref{n-p(t)-1}) the asymptotic form (\ref{MathcalApm1}) of the modes in the future infinity, where the coefficients $\mathcal{A}_\pm(\mathbf{p})$ are defined in (\ref{MathcalApm}). A similar expression one also obtains for the anomalous quantum expectation values. 

There are two comments in order at this point. First, one can see that the secular (growing with time of observation, $t$, and/or with the duration of the pulse, $T$) contributions that we obtain in the background fields are essentially due to the presence of non-trivial collision integrals in strong background fields. Namely, in the strong background field under consideration there is no energy conservation and, hence, various processes, such as creation of charged particles together with photons from the initial state of the theory, are allowed. Such processes are forbidden without background fields. In fact, the growing with time level population from (\ref{n-p(t)-1}) is just the collision integral integrated over the time of the evolution of the system. The same is true for (\ref{2-loop-n_p(T)}) and (\ref{2-loop-kappa_p(T)}). 

Second, as we have already discussed, after the pulse is turned off the time change of the level population and anomalous average is a set up for a different problem. In fact, during the duration of the background electric pulse there can be generated in principle some level population and anomalous averages. The latter will set up an initial data for the thermalization process after the switching off of the background field. Such a process in a bit different theory was discussed in \cite{Akhmedov:2021vfs}. During this thermalization process, which is just a redistribution of the level populations and anomalous averages for different levels in a situation without a background field, there can appear an electric current, but in this paper we are not discussing this problem, as we have already explained above.

Thus, we concentrate on the situation within the duration of the background pulse, i.e. on the time period $ |t|<T $. The integration over the region $ \mathcal{T}<-T $ in (\ref{n-p(t)-1}), i.e. before the start of the pulse, is suppressed, since all expectation values are taken with respect to the standard Poincare invariant state (the in-state before the pulse). At the same time, as we have explained above, within the pulse duration we can approximate the exact modes with the parabolic cylinder functions. Then, for the level population we obtain the following expression:
\begin{equation}
\begin{split}
n_\mathbf{p}(T)\simeq \mathcal{N}_{\mathbf{p}_\perp}(\chi) &= -2e^5E
\int\limits_{\chi-2\sqrt{eE}T}^{\chi}dX
\int\limits_{-\infty}^{+\infty}d\widetilde{\chi}
\int\dfrac{d^3\mathbf{Q}_1d^3\mathbf{Q}_2}{(2\pi)^6}
\dfrac{e^{i(|\mathbf{Q}_1|+|\mathbf{Q}_2|)\widetilde{\chi}}}
{|\mathbf{Q}_1||\mathbf{Q}_2|}\times \\
&\times
g^*_{\mathbf{p}_\perp}(X+\widetilde{\chi}/2)
g^*_{\mathbf{p}_\perp-\mathbf{q}_{1\perp}-\mathbf{q}_{2\perp}}(X+\widetilde{\chi}/2-Q_1-Q_2)\times \\
&\times
g_{\mathbf{p}_\perp-\mathbf{q}_{1\perp}-\mathbf{q}_{2\perp}}(X-\widetilde{\chi}/2-Q_1-Q_2)
g_{\mathbf{p}_\perp}(X-\widetilde{\chi}/2),
\end{split}
\end{equation}
where we have performed the change of variables (\ref{dimensionless_vars}).

As in the case of the one-loop correction to the current \cite{Akhmedov:2023zfy}, we can extend the integration over $ X $ onto the entire real axis with accuracy $ \mathcal{O}(1/T) $ and obtain that $ n_\mathbf{p}(T) $ does not depend on $ T $ in the leading order. Thus, the corresponding correction to the current also does not give an additional power of growth with $ T $. For the anomalous average $ \kappa_\mathbf{p}(T) $, the situation is similar. Thus, the diagram shown in fig. \ref{image-two-loop}, being suppressed by the extra power of the fine structure constant, does not provide a correction which grows with $T$ faster than the tree-level contribution to the current.

\subsubsection{Shifted sunset diagram}\label{Section_Shifted}

\begin{figure}[h]
\center{\includegraphics[width=0.8\linewidth]{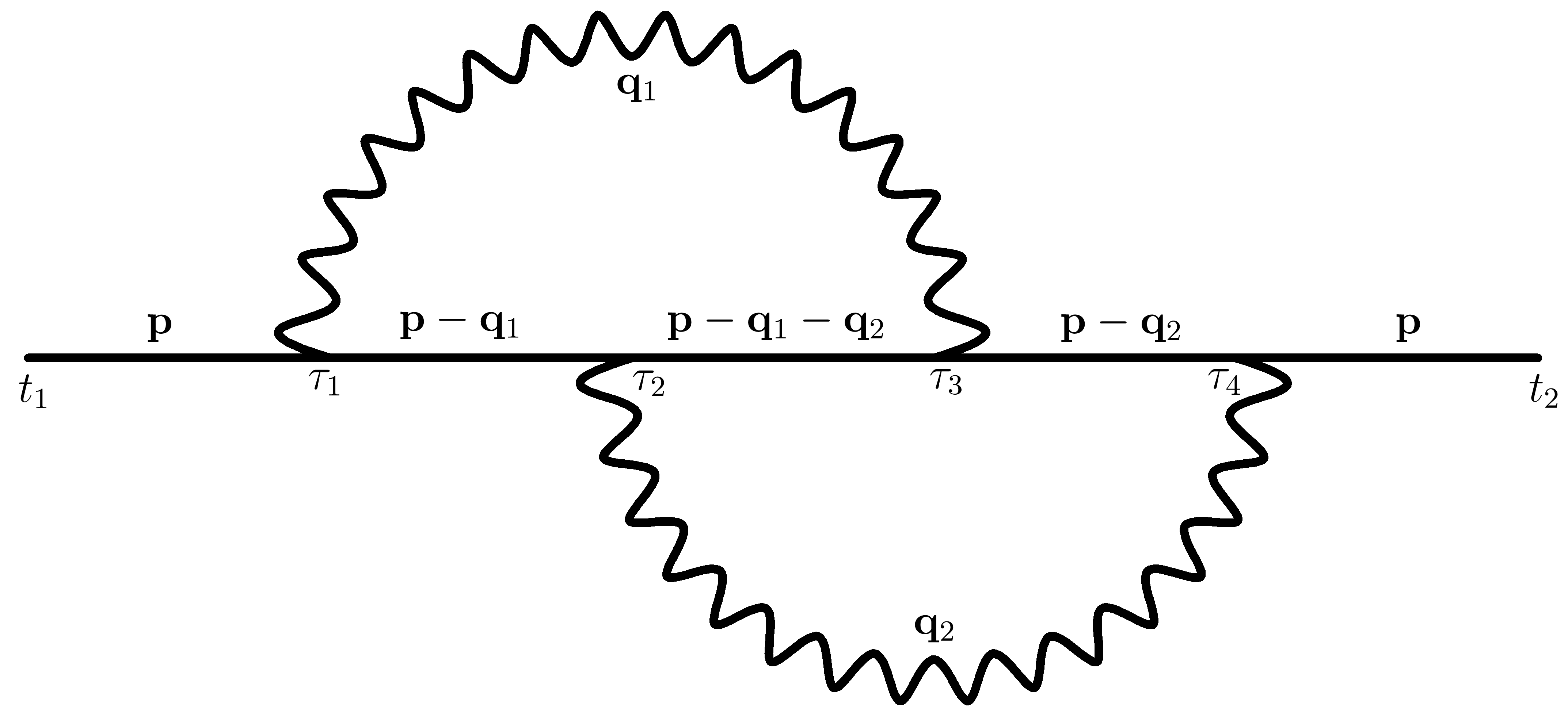}}
\caption{Shifted sunset diagram}
\label{image_dia_shifted_sunset}
\end{figure}

Consider now the diagram from the Fig. \ref{image_dia_shifted_sunset}.
The corresponding general expression for the correction to the propagator can be written as:
\begin{equation}\label{Delta_D_ab}
\begin{split}
\Delta D^{ab}(\mathbf{p};t_1,t_2)=e^4\sum\limits_{cdef}\sgn(c)\cdots\sgn(f)\int d\tau_1\cdots d\tau_4\int\dfrac{d^3\mathbf{q}_1 d^3\mathbf{q}_2}{(2\pi)^6}\, G^{ce}(\mathbf{q}_1;\tau_1,\tau_3)
G^{df}(\mathbf{q}_2;\tau_2,\tau_4)\times \\
\times \mathbb{D}\left[
D^{ac}(\mathbf{p};t_1,\tau_1)D^{cd}(\mathbf{p}-\mathbf{q}_1;\tau_1,\tau_2)
D^{de}(\mathbf{p}-\mathbf{q}_1-\mathbf{q}_2;\tau_2,\tau_3)
D^{ef}(\mathbf{p}-\mathbf{q}_2;\tau_3,\tau_4)
D^{fb}(\mathbf{p};\tau_4,t_2)\right],
\end{split}
\end{equation}
where the indexes, $a,b$ and etc., in this expression are the same as in (\ref{loop_finally_dia}) and
\begin{equation}
\mathbb{D}^{\mu\nu\alpha\beta}=
\left(\left[D^\mu_1\right]^\dagger-D^\mu_1\right)
\left(\left[D^\nu_2\right]^\dagger-D^\nu_2\right)
\left(\left[D^\alpha_3\right]^\dagger-D^\alpha_3\right)
\left(\left[D^\beta_4\right]^\dagger-D^\beta_4\right),
\end{equation}
while $ \mathbb{D}=\mathbb{D}^{\mu\nu\alpha\beta}g_{\mu\alpha}g_{\nu\beta} $.

To represent this expression and the corresponding correction to the current in terms of modes we rewrite the tree-level propagators in the following form:

\begin{equation}\label{A&B}
D^{ab}(\mathbf{p};t_1,t_2)=A^{ab}(\mathbf{p};t_1,t_2)+B^{ab}(\mathbf{p};t_1,t_2),
\end{equation}
where
\begin{equation}\label{A&B_exact_form}
\begin{split}
A^{ab}(\mathbf{p};t_1,t_2) &= \begin{pmatrix}
\theta(t_1-t_2) & 0 \\
1 & \theta(t_2-t_1)
\end{pmatrix}f_\mathbf{p}(t_1)f^*_\mathbf{p}(t_2), \\
B^{ab}(\mathbf{p};t_1,t_2) &= \begin{pmatrix}
\theta(t_2-t_1) & 1 \\
0 & \theta(t_1-t_2)
\end{pmatrix}f^*_\mathbf{p}(t_1)f_\mathbf{p}(t_2),
\end{split}
\end{equation}
and similarly for the photon propagator:
\begin{equation}
G^{ab}(\mathbf{p};t_1,t_2)=\mathcal{A}^{ab}(\mathbf{p};t_1,t_2)+\mathcal{B}^{ab}(\mathbf{p};t_1,t_2).
\end{equation}
To simplify the expressions below let us introduce the following notation $ A^{ab}(\mathbf{p};t_1,t_2)=A^{ab}_{12} $. Furthermore, we also use the following notations: $ A^{ab}(\mathbf{p};t,\tau_1)=A^{ab}_{01} $ etc. Then, one can restore the expanded expression (with all the arguments explicitly written) unambiguously. Hence, e.g.

\begin{equation}\label{D-D-1}
D^{+c}(\mathbf{p};t,\tau_1)D^{f-}(\mathbf{p};\tau_4,t)=A^{+c}_{01}A^{f-}_{40}+B^{+c}_{01}B^{f-}_{40}+A^{+c}_{01}B^{f-}_{40}+B^{+c}_{01}A^{f-}_{40}.
\end{equation}
The first two terms from the right side of (\ref{D-D-1}) are proportional to $ |f_\mathbf{p}(t)|^2 $ and, hence, contribute to $ n_\mathbf{p}(t) $. The third term is proportional to $ f^2_\mathbf{p}(t) $ and, thus, contributes to $ \kappa^*_\mathbf{p}(t) $, as can be seen from (\ref{exact_current_finally}). The last term contributes to $ \kappa_\mathbf{p}(t) $.

After the substitution of (\ref{A&B}) into (\ref{Delta_D_ab}), we obtain a sum of integrals over $ \tau_{1,2,3,4} $, whose integration domains are specified by inequalities of the form $ \tau_i<\tau_j $, for some $ i,j $. However, from the causality of the Schwinger-Keldysh technique (see e.g. \cite{Akhmedov:2022uug}), we know for sure that all $ \tau_i $ do not exceed $ t $: otherwise the loop contribution is vanishing.

Since the product $ G^{ce}(\mathbf{q}_1;\tau_1,\tau_3)
G^{df}(\mathbf{q}_2;\tau_2,\tau_4) $ contains terms proportional to the product of exponents $ e^{\pm i|\mathbf{q}_1|(\tau_1-\tau_3)}e^{\pm i|\mathbf{q}_2|(\tau_2-\tau_4)} $, after the change of variables,
\begin{equation}\label{new_times_2.0}
\mathcal{T}_1=\dfrac{\tau_1+\tau_3}{2},\;\mathcal{T}_2=\dfrac{\tau_2+\tau_4}{2};\quad \widetilde{\tau}_1=\tau_1-\tau_3,\;\widetilde{\tau}_2=\tau_2-\tau_4,
\end{equation}
the integral over $ \mathbf{q}_{1,2} $ is saturated in the region $ |\widetilde{\tau}_{1,2}|\ll |\mathcal{T}_{1,2}| $ (for a large $ \widetilde{\tau_i} $, the exponent $ e^{\pm 2i |\mathbf{q}_i|\widetilde{\tau_i}} $ is fast oscillating, and the integral over $ \mathbf{q}_i $ is suppressed) and we can extend the integration over $ \widetilde{\tau}_{1,2} $ onto the entire real axis or the real semi-axis (some terms can contain theta-functions, such as $ \theta(\tau_1-\tau_3) $ etc.). In any of these cases, after such an extension, the integrals over $ \widetilde{\tau}_i $ no longer contain $ t $ within the limits of integration.

As was explained above, we are interested only in the integration regions inside the pulse,$ -T<\mathcal{T}_i<t<T $. In such a case we can use the parabolic cylinder functions (\ref{straight_modes}) as the approximate form of the modes. Below we will use that the coefficient in (\ref{straight_modes}) obeys:
\begin{equation}\label{mod_c}
\left|c(\mathbf{p}_\perp)\right|^2=\dfrac{1}{\sqrt{2eE}}\exp\left[-\dfrac{\pi(\mathbf{p}^2_\perp+m^2)}{4eE}\right].
\end{equation}
One can see from (\ref{A&B}) and (\ref{A&B_exact_form}) that $ D^{a_1 a_2}(\mathbf{k},t_i,t_j)\propto |c(\mathbf{k}_\perp)|^2 $ and does not depend on $ T $ in the leading order. Thus, all the dependence on $ T $ in (\ref{Delta_D_ab}) is contained only in the limits of integration over $ \mathcal{T}_i $.

For each integral over $ \mathcal{T}_1 $ one has the integrand of the form
\begin{equation}\label{ffff_2_loop}
\begin{aligned}
\left(\left[D^\mu_1\right]^\dagger-D^\mu_1\right)
\left(\left[D_{3,\mu}\right]^\dagger-D_{3,\mu}\right) &
\Big{[}g_{\mathbf{p}_\perp}(p_1+eE\tau_1)          
g_{\mathbf{p}_\perp-\mathbf{q}_{1,\perp}}(p_1-q_1+eE\tau_1) \times \\
&\times
g_{\mathbf{p}_\perp-\mathbf{q}_{1,\perp}-\mathbf{q}_{2,\perp}}
(p_1-q_1-q_2+eE\tau_3) 
g_{\mathbf{p}_\perp-\mathbf{q}_{2,\perp}}(p_1-q_2+eE\tau_3)\Big{]},    
\end{aligned}
\end{equation}
where some modes $ g $ can contain the sign of the complex conjugation. To single out the leading contribution  for large $ T $ from the integrals under consideration we ought to use the asymptotic behavior of the parabolic cylinder functions with a large argument. Regardless of whether the mode is complex conjugate or not, it has the asymptotic behavior of the form (\ref{Cylinder_2}). Consequently, different contributions will differ only in the coefficients, which are products of $ \alpha $'s and $ \beta $'s and their arguments. Meanwhile, the operator 

$$ \left(\left[D^\mu_1\right]^\dagger-D^\mu_1\right)\left(\left[D_{3,\mu}\right]^\dagger-D_{3,\mu}\right) $$ 
acts on the asymptotic form (\ref{Cylinder_2}) as follows:
\begin{equation}
\left(\left[D^\mu_1\right]^\dagger-D^\mu_1\right)
\left(\left[D_{3,\mu}\right]^\dagger-D_{3,\mu}\right)\mapsto
(2\mathbf{p}_\perp-\mathbf{q}_{1,\perp})
(2\mathbf{p}_\perp-2\mathbf{q}_{2,\perp}-\mathbf{q}_{1,\perp}).
\end{equation}
In this case such an integrand as (\ref{ffff_2_loop}) is of the order $ 1/\mathcal{T}_1^2 $, which means that the integral over $ \mathcal{T}_1 $ converges at infinity and does not grow with $ T $. Up to the order of $ \mathcal{O}(1/T) $, one can extend the integration over $ \mathcal{T}_1 $ to the entire real axis. For the integral over $ \mathcal{T}_2 $, the situation is similar. Thus, in the case of the diagram under consideration, the quantum corrections again do not give an additional growth with $ T $, which means that the final contribution to the current will be just suppressed in comparison with the tree-level current by the extra power of the fine structure constant (and will not contain any higher power of $T$). In all, we find that the diagram shown in Fig. \ref{image_dia_shifted_sunset} is suppressed for large values of $ T $.

\subsubsection{Two-loop tadpole diagram}

\begin{figure}[h]
\center{\includegraphics[width=0.5\linewidth]{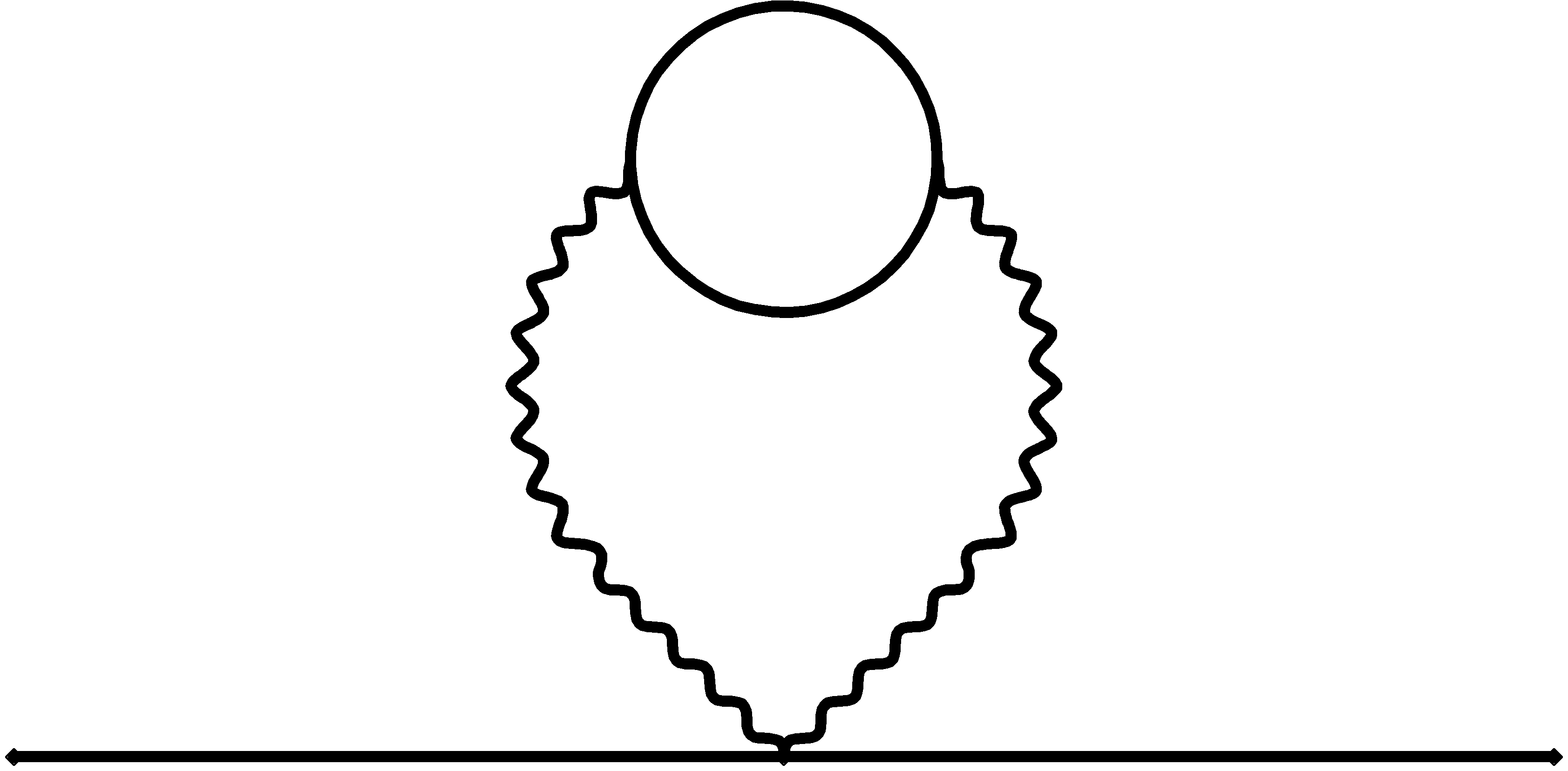}}
\caption{Two-loop tadpole diagram. In the Schwinger-Keldysh technique there is a sum af such diagrams}
\label{image_dia_tadpole_double}
\end{figure}

In this subsection we consider the contribution of the two-loop tadpole type diagram shown on the Fig. \ref{image_dia_tadpole_double}.
The general expression for the correction to the matrix of propagator corresponding to this diagram is as follows:
\begin{equation}\label{Nonameeq}
\Delta D^{ab}(\mathbf{p};t_1,t_2)=ie^2\sum\limits_{c}\sgn(c)
\int d\tau
\int\dfrac{d^3\mathbf{q}}{(2\pi)^3} \Delta G^{cc}(\mathbf{q};\tau,\tau)
D^{ac}(\mathbf{p};t_1,\tau)D^{cb}(\mathbf{p};\tau,t_2),
\end{equation}
where $ \Delta G^{cc}(\mathbf{q};\tau,\tau) $ is the one-loop corrected photon propagator taken at coinciding times. It was calculated in \cite{Akhmedov:2023zfy}. The expressions for $ \Delta G^{A,R}(\mathbf{q};\tau_1,\tau_2) $ have the same retarded (advanced) structure as the corresponding tree-level propagators. In particular, $\Delta G^{A,R}(\mathbf{q};\tau,\tau)=0 $ for all $ \tau $. In this case, we obtain that
\begin{equation}
 \Delta G^{cc}(\mathbf{q};\tau,\tau) = \Delta G^{K}(\mathbf{q};\tau,\tau) 
\end{equation}
for all $ c=\pm 1 $. Furthermore, the integral over $ \mathbf{q} $ in (\ref{Nonameeq}) is factorizable, as in the case of the one-loop tadpole correction. However, in this case the factorized integral depends on $ \tau $. Namely, the leading contribution is as follows:
\begin{equation}\label{massshift}
\int\dfrac{d^3\mathbf{q}}{(2\pi)^3}\Delta G^{K}(\mathbf{q};\tau,\tau)=
\int\dfrac{d^3\mathbf{q}}{(2\pi)^3}\dfrac{n^\mu_\mu(\mathbf{q},\tau)}{|\mathbf{q}|}
= e^2(\tau+T)\mathcal{C}',
\end{equation}
where $ \mathcal{C}' $ is some constant similar $ \mathcal{C} $ from the one-loop tadpole diagram which was calculated in \cite{Akhmedov:2023zfy}. Thus, in contrast to the one-loop tadpole diagram, the correction to the current contains an additional dependence on $ \tau $ in its integrand:
\begin{equation}
\Delta j_1^{(\mathrm{tad})}(t)=2 e \int \frac{d^3 \mathbf{p}}{(2 \pi)^3}\left(p_1+e E T\right)\left[f_{\mathbf{p}}^2(t) \chi_{\mathbf{p}}(t)+\text { h.c. }\right],
\end{equation}
where
\begin{equation}\label{tadpole-2-lopp-kappa}
\chi_{\mathbf{p}}(t)=i e^4 \mathcal{C}' \int\limits_{t_0}^t
(\tau+T)\left[f_{\mathbf{p}}^*(\tau)\right]^2 d \tau.
\end{equation}
However, similarly to the one-loop tadpole corrections, one can show that the integral over $ \tau $ in $ \Delta j_1^{(\mathrm{tad})} $ can grow with $ T $ only for $ \tau\in[-T,T] $, i.e. during the pulse. One also can show, using the asymptotics (\ref{Cylinder_2}), that the sum of the integrals over $ \tau $ in $ \Delta j_1^{(\mathrm{tad})} $ (the integral in (\ref{tadpole-2-lopp-kappa}) and its h.c.) converges and is saturated from the vicinity of such $\tau$ that $ \sqrt{eE}\tau\sim 1 $, for which $ (T+\tau)\simeq T$. Taking this into account, one can see that all calculations in this case are the same as for the one-loop tadpole correction. Thus, the correction under consideration can be absorbed by just a shift of the mass by the quantity $ \mathcal{C}_\text{2-loop}=e^2 T\mathcal{C}' $, following from (\ref{massshift}), rather than attributed to a growth of the current.

\subsubsection{Taking into account the $ \lambda|\phi|^4 $ self-interaction}

\begin{figure}[h]
\center{\includegraphics[width=0.7\linewidth]{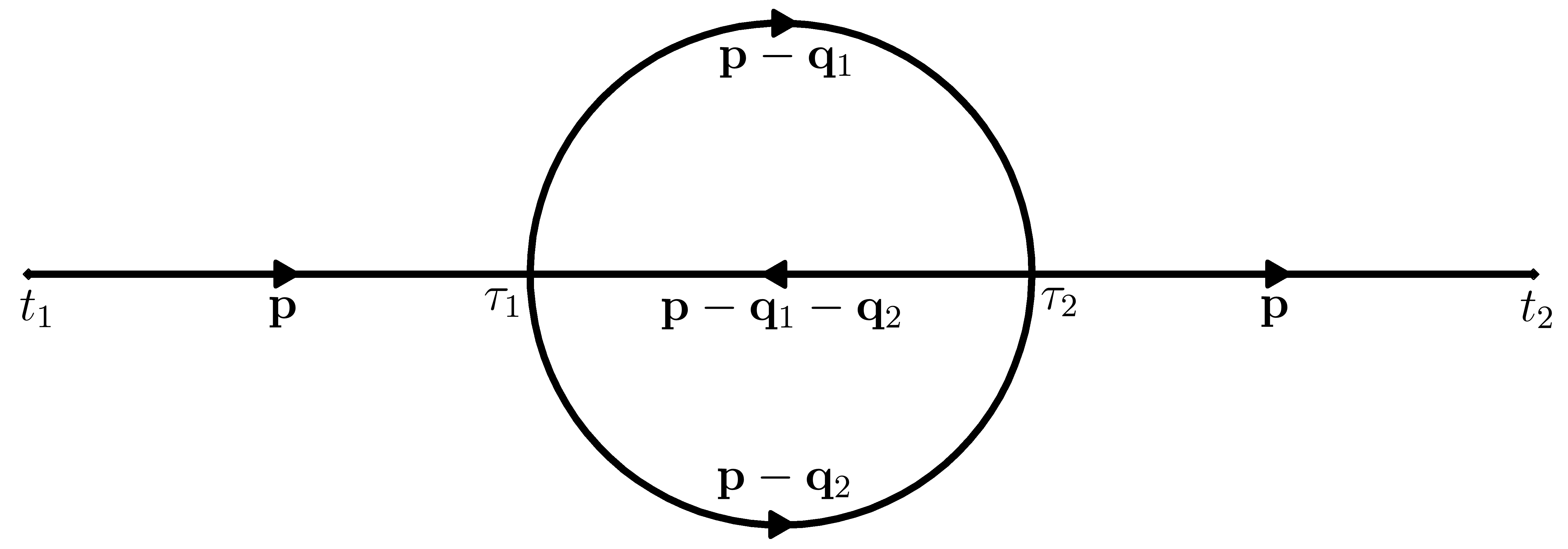}}
\caption{Sunset diagram for the $ \lambda|\phi|^4 $ theory}
\label{image_dia_lambda_sunset}
\end{figure}

In principle, the scalar electrodynamics can also contain an additional self-interaction terms, such as e.g. $ \lambda|\phi|^4 $. Then the action of the theory is:
\begin{equation}\label{SQED_action_finally}
S[\phi,\phi^\dagger; a^\mu]=\int d^4x\left[|D_\mu\phi + iea_{\mu}\phi|^2-m^2|\phi|^2-\dfrac{1}{4}f_{\mu\nu}f^{\mu\nu}+
\dfrac{\lambda}{4!}|\phi|^4\right].
\end{equation}
In such a case, there is another two-loop diagram which may lead to a secular contribution to the current. This diagram is shown on the fig. \ref{image_dia_lambda_sunset}. (One loop tadpole correction to the Keldysh propagator for the $ \lambda|\phi|^4 $ theory can be shown also to give a shift the mass in the theory.)
\\

The general expression for the correction to the matrix of propagators following from this diagram is as follows:
\begin{equation}\label{Delta_D_ab_lambda}
\begin{split}
\Delta D^{ab}(\mathbf{p};t_1,t_2)=\lambda^2\sum\limits_{cd}\sgn(c)\sgn(d)\int d\tau_1 d\tau_2\int\dfrac{d^3\mathbf{q}_1 d^3\mathbf{q}_2}{(2\pi)^6}\times \\
\times
D^{ac}(\mathbf{p};t_1,\tau_1)D^{cd}(\mathbf{p}-\mathbf{q}_1;\tau_1,\tau_2)
D^{dc}(\mathbf{p}-\mathbf{q}_1-\mathbf{q}_2;\tau_2,\tau_1)
D^{cd}(\mathbf{p}-\mathbf{q}_2;\tau_1,\tau_2)
D^{db}(\mathbf{p};\tau_2,t_2).
\end{split}
\end{equation}
The dependence on time in the expression (\ref{Delta_D_ab_lambda}) is the same as of the other sunset diagrams, which we have considered above. Namely, the limits of integration over $ \tau_{1,2} $ for $ n_\mathbf{p}(t) $ and $ \kappa_\mathbf{p}(t) $ are the same as in the expressions (\ref{n-p(t)}) and (\ref{kappa-p(t)}) correspondingly. Then, when $\tau_{1,2}$ are in the region before the pulse is turned on the situation is trivial, i.e. the same as in the absence of any external field. After the pulse is turned off, via a Bogolyubov transformation one can again reduce everything down to the situation without the external field, but with the initial quantum expectation values being not equal to zero. It is known that in such a situation there is a secular growth with $ t $ associated with the process of thermalization similar to the one considered in \cite{Akhmedov:2021vfs}, as we have mentioned above.

Let us now concentrate on the situation within the duration of the external pulse, $ |\tau_{1,2}|<T $. One can notice that the diagram from the Fig. \ref{image_dia_shifted_sunset} can be reduced to the diagram from the Fig. \ref{image_dia_lambda_sunset} if one shrinks photon propagators into points. As a result, to obtain (\ref{Delta_D_ab_lambda}) from (\ref{Delta_D_ab}) one should perform the following changes: 
\begin{equation}
\sgn(e)G^{ce}(\mathbf{q}_1;\tau_1,\tau_3)\mapsto \delta_{ce}\delta(\tau_1-\tau_3),\quad
\sgn(f)G^{df}(\mathbf{q}_2;\tau_2,\tau_4)\mapsto \delta_{df}\delta(\tau_2-\tau_4),\quad
e^4 \mathbb{D}\mapsto \lambda^2.
\end{equation}
This means that the integrals over $ \tau_{1,2} $ in (\ref{Delta_D_ab_lambda}) are very similar to the integrals over $ \mathcal{T}_{1,2} $ in considered in the subsection \ref{Section_Shifted}: using the asymptotic form (\ref{Cylinder_2}) we find that the integrals over $ \tau_i $ also have the integrands of the order $ 1/\tau_i^2 $ and do not grow with $ T $.

In all, we find that taking into account the self-interaction of the scalar field does not lead to an additional growth of the electric current with $T$, in comparison with the tree-level result.

\section{Discussion and Conclusion}

We have shown that at the second loop order the only contribution to the current, that grows with the pulse duration $T$ fast enough ($\sim T^2$ or faster) to substantially change the tree-level current (which grows linearly with $T$), follows from the diagrams depicted on the Fig. \ref{image_dia_photon}. Other diagrams from Fig. \ref{image-two-loop}, \ref{image_dia_shifted_sunset}, \ref{image_dia_tadpole_double} and 
\ref{image_dia_lambda_sunset} lead either to the mass renormalization or suppressed in the limit of large $T$. (They, however, may contain contributions growing with $t$, which are gained after the switching off of the pulse.) Therefore, a strong modification of the tree-level Schwinger's current does occur at the two-loop level and is due to the creation of photons, which in their own right decay into pairs.

Thus, the main conclusion of the present paper: for the lengthy enough background pulse the Schwinger's expression for the current of the created particles is drastically modified. This, in particular, means that for long enough pulse one has to perform a resummation of the leading contributions from all loop orders. The form of the leading contributions may depend on whether the background field is weak, $ eE\ll m^2 $, or strong, $ eE\geqslant m^2 $. We will perform the corresponding calculations in our forthcoming papers.

This work was supported by Russian Science Foundation (Project Number: 23-22-00145).

\bibliography{bibliography}
\bibliographystyle{unsrt}
\end{document}